# Autonomous collision attack on OCSP services

Ken Ivanov
EldoS Corporation

Revision 1.0 (August 2016)

## Abstract

The paper describes two important design flaws in Online Certificate Status Protocol (OCSP), a protocol widely used in PKI environments for managing digital certificates' credibility in real time. The flaws significantly reduce the security capabilities of the protocol, and can be exploited by a malicious third party to generate forged signed certificate statuses and, in the worst scenario, forged certificates. Description of the flaws, along with expected exploitation routes, consequences for consuming application layer protocols, and proposed countermeasures, is given.

## 1 Introduction

Online certificate status protocol (OCSP) [1] is one of the two common methods of obtaining up-to-date digital certificate status information from certification authorities (CA) in modern public key infrastructures (PKI) (the other is based on certificate revocation lists [2]). An entity willing to obtain the latest update on a certificate status – such as a web browser or a program that needs to establish the validity of a digital signature – connects to a dedicated web service maintained by the CA or its affiliated party, called an OCSP responder, and sends in a request, including an identifier of the certificate in question. The OCSP responder then provides the requestor with a certificate status record, which includes the certificate identifier, its up-to-date status as maintained by the CA ('active', 'revoked' or 'unknown'), the time of the last status information update, and when to expect the next update. The status record, called an OCSP response, is digitally signed with the OCSP responder's certificate, which certifies its authenticity on behalf of the CA and makes it a standalone verifiable PKI entity.

OCSP-based validation schemes are widely used in modern PKI infrastructures, particularly with advanced electronic signatures (CAdES, PAdES, XAdES and ASiC) [3]. Typical OCSP usage scenarios include straightforward synchronous real-time checks for up-to-date certificate status, e.g. when authenticating a TLS server, and retrieval of OCSP certificate status records for insertion into long-term electronic signatures.

## 2 Terms

While we don't expect the reader to be an expert in PKI, familiarity with basic principles and practical implementations of X.509-based public key environments will make it easier to understand the impacts the exploited vulnerabilities may have on the environments. In particular, knowledge and understanding of such concepts as certificates, certification authorities, world of trust, authentication, revocation, and digital signatures would contribute much towards understanding the paper. Below we provide definitions for the main concepts used in the paper; please note that the definitions are not exhaustive, and only cover the aspects of the corresponding essences relevant to the discussed issues for the sake of simplicity.

*Digital certificate (or simply certificate)*: an electronic document that binds information of a physical or electronic entity to its cryptographic public key, and certifies it with a digital signature of an authorized higher-level entity (certification authority). The matching private key remains in the possession of the entity owning the certificate and, unlike the certificate, is never disclosed to third parties. A digital certificate typically includes the name or ID of the owner, the purposes for the public key it carries (digital signing, key exchange, certificate signing, etc.), and its validity period.

*Certification authority (CA)*: a designated entity having a right of issuing digital certificates to other parties by signing them with its private key. CA is identified by its own digital certificate, which is used by third parties to validate signatures made by the CA's private key over the certificates it had issued.

*Certificate status*: a certificate validity indicator from the authorization perspective (typical values: active, revoked, and unknown). Certificate statuses are maintained and updated by the CA, and are made available to third parties via its revocation services.

*Certificate chain*: a sequence, or sometimes an inverted tree, of certificates where each certificate except the first one is a CA for the certificate immediately preceding it. The first certificate in the sequence is an end-entity certificate used for application purposes, and the last certificate is typically a root certificate and a trust anchor (unconditionally trusted).

*Public key infrastructure*: a framework consisting of a set of roles, policies, and procedures set up to manage trust relationships in multi-user environments, and based on features offered by asymmetric cryptography to enforce dependency and trust links between its members.

*Certificate validation*: establishing the fact that a particular certificate has authority to create particular kinds of signatures; normally done by walking up the certificate chain and validating each link constituting it. A validation of a typical application layer signature involves validation of a number of certificates (implying chains): the signing end-entity certificate itself, a number of OCSP and CRL issuers' certificates, TSA certificates, and so on.

*Revocation service*: a publicly available service provided by a CA which provides means to establish the up-to-date status of certificates issued by that CA. If a particular certificate is reported to the CA as having been compromised, the CA changes its status to 'revoked' and propagates the updated information via its revocation services, typically web-based CRL and OCSP endpoints.

*Online Certificate Status Protocol (OCSP)*: a real-time request-response lightweight protocol allowing third parties to request an up-to-date status for a particular certificate from the CA.

*OCSP requestor*: a party of the OCSP protocol which initiates the protocol by sending a certificate status request to the other party, OCSP responder. This is normally an entity willing to establish validity of the certificate, and needing to make sure that the certificate is not revoked.

*OCSP responder*: a party of the OCSP protocol which serves incoming certificate status requests, and issues the corresponding responses containing the up-to-date certificate status together with some additional information, and signed with the responder's private key. OCSP responders are normally set up and maintained by the CA or an entity affiliated with it.

## 3 OCSP: the protocol and its applications

### 3.1 Protocol details

**Protocol overview.** OCSP is a lightweight stateless one-step request-response protocol which typically works over HTTP(S) transport. A party which needs to ensure that a certificate is legitimate and had not been revoked by its issuer establishes the location of the relevant OCSP responder and posts a status request for the certificate in question. The certificate is identified in the request by its serial number, which is unique within the same CA. Besides the serial number, the request includes the details of the CA certificate, and may optionally include a few extension fields, such as a random nonce. A few OCSP responders also require the requests to be signed by the requestors. Enabling this option has no effect on the vulnerability discussed in this paper.

Upon receiving the request, the responder performs a check for the requested certificate in its local database and retrieves the most recent status it has for the certificate. It then creates an authorized response by combining together the certificate identifier, the status flag, the time information (the last known status update and the next expected status update times, and response creation time), and signing the created record with its authorized OCSP certificate.

An important point here is that both the request and the response identify the certificate by its serial number. Neither party needs to possess any knowledge about the certificate contents to be able to compose valid requests and responses. This detail makes a significant contribution to the exploitation of the described vulnerability.

The figures below show the formats of OCSP request and response as defined in [1]. The protocol employs ASN.1 format [4], and uses DER encoding rules [5] for all messages.

```
OCSPRequest ::= SEQUENCE {
    tbsRequest              SEQUENCE {
        version             [0]     EXPLICIT Version DEFAULT v1,
        requestorName       [1]     EXPLICIT GeneralName OPTIONAL,
        requestList                 SEQUENCE OF SEQUENCE {
            reqCert                 SEQUENCE {
                hashAlgorithm       AlgorithmIdentifier,
                issuerNameHash      OCTET STRING, -- Hash of issuer's DN
                issuerKeyHash       OCTET STRING, -- Hash of issuer's public key
                serialNumber        CertificateSerialNumber
            },
            singleRequestExtensions    [0] EXPLICIT Extensions OPTIONAL
        },
        requestExtensions   [2]     EXPLICIT Extensions OPTIONAL
    },
    optionalSignature       [0]     EXPLICIT Signature OPTIONAL
}
```

**Fig. 1.** OCSP request format, as sent by the requestor to the OCSP responder.

```
BasicOCSPResponse ::= SEQUENCE {
    tbsResponseData     SEQUENCE {
        version             [0] EXPLICIT Version DEFAULT v1,
        responderID             ResponderID,
        producedAt              GeneralizedTime,
        responses               SEQUENCE OF SEQUENCE {
            certID              CertID,
            certStatus          CertStatus,
            thisUpdate          GeneralizedTime,
            nextUpdate          [0]     EXPLICIT GeneralizedTime OPTIONAL,
            singleExtensions    [1]     EXPLICIT Extensions OPTIONAL
        },
        responseExtensions   [1] EXPLICIT Extensions OPTIONAL
    },
    signatureAlgorithm  AlgorithmIdentifier,
    signature           BIT STRING,
    certs           [0] EXPLICIT SEQUENCE OF Certificate OPTIONAL

}
```

**Fig. 2.** OCSP response format, as sent by the OCSP responder back to the requestor.

**OCSP certificate.** The location(s) of the OCSP responder that provides status information for a particular certificate is included in the certificate by its issuing CA in the form of Authority Information Access extension, and is a non-detachable compound of the certificate. Infrastructure-wise, the position of the OCSP responder's certificate in the PKI interrelationship graph is strictly defined (Fig. 3). The certificate that signs OCSP responses either must be the same certificate that issued the certificate being checked (i.e. the CA certificate), or it must be issued by the CA certificate specifically for OCSP signing purposes, which is indicated in an Extended Key Usage extension in the designated certificate. An option of explicitly trusted OCSP responder is also available, but it does not agree well with PKI principles, and is rarely used in practice.

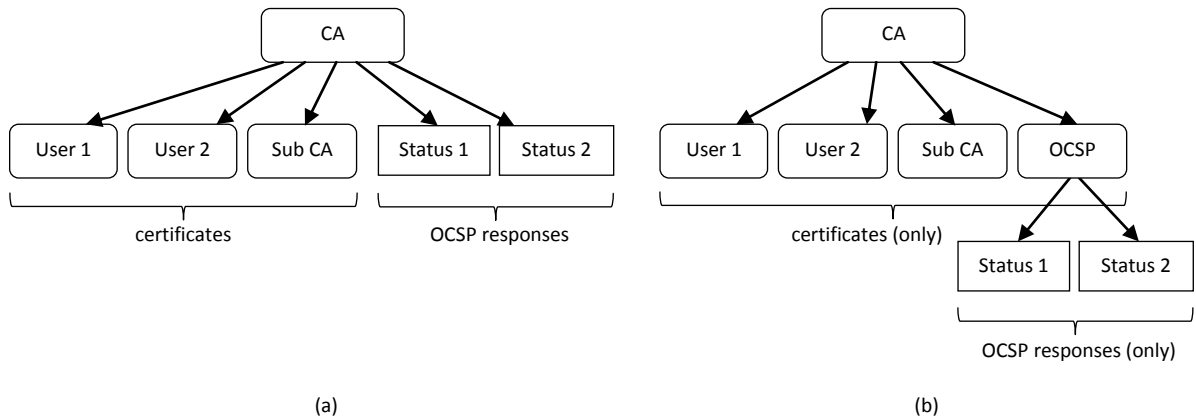

**Fig. 3**. Two options of positioning the OCSP responder in a PKI tree; (a) OCSP statuses are issued by the CA; (b) dedicated OCSP certificate.

**Nonce extension.** [1] suggests requestors to address replay attacks by including random pieces of data with all requests via the protocol Nonce extension. The responder mirrors back the nonce by including it in its signed response. The requestor should reject the response if the nonce value it contains doesn't match the original. Support and specifics of handling of the nonce extension by implementations play an important role in the described vulnerability.

**Lightweight profile.** Having been originally designed as a real-time protocol capable of providing up-to-date / most recent information on certificate status, OCSP and its procedures were subject to a significant conceptual change later. Back in 2007, following a joint effort from Verisign and Microsoft, a special procedure called a 'Lightweight OCSP profile' [6], was introduced to help high-volume PKI environments cope with burden caused by a flood of OCSP requests on their OCSP responders. The main idea of the procedure was to refuse from generation of OCSP responses in real time in favour of pre-calculated responses, to decrease service times and improve fault tolerance of OCSP responders. This approach obviously brings in a certain level of latency to reporting actual certificate status, and effectively puts OCSP mechanisms closer to CRL-based scenarios. We are not going to discuss the applicability and effectiveness of the lightweight profile in this paper, but as this approach is used widely by mass service CAs, it is useful to know it exists.

**Non-existent serial number handling.** OCSP specification introduces certain rules for the OCSP responders regarding handling of non-existent certificate serial numbers, should they be specified in the request. A compliant responder should come up with the 'unknown' status for such serial numbers. In practice, however, different approaches are also used, with some responders simply closing the connection without providing any reasonable response, and others providing Good status. The latter is probably due to clarification given in [1, p.2.2] for Good status, which provides that this status can be returned if the certificate was never issued. An alternative assumption is that the responders simply browse their revoked certificates database for the requested serial number, and return Good status if no revoked certificate with the provided number was found.

**Time values accuracy.** OCSP specification does not provide any guidance on accuracy of time values included with the response – 'this update', 'next update' and 'produced at'. In practice most of responders stick to the accuracy of one second.

### 3.2 Usage scenarios

OCSP is widely used in modern PKI environments. Its main purpose is providing a reliable, authenticated and authorized indicator of the certificate status at the given moment in time. As such, OCSP-based schemes are used in a variety of higher-level protocols.

**Digital signatures.** The most straightforward use of OCSP is as a compound of the signing certificate chain validation routine performed within the process of validating generic digital signatures (PKCS#7-compliant and similar). In this scenario a set of OCSP responders are contacted by the signature verifier to establish the validity of the signing certificate and certificates comprising its chain at the moment of signature validation. A number of widely used application-layer protocols and applications, including PDF signatures [7], XML-SIG [8], and S/MIME [9], rely on generic PKCS#7 signatures and associated OCSP services internally.

**Advanced electronic signatures.** Within Advanced Electronic Signature infrastructures (AdES) framework signed OCSP responses are embedded into the body of the signature together with the rest of validation information (certificates and CRLs), effectively converting the signature to a standalone self-contained signed entity which can be validated without referring to online validation information sources. A special case of AdES signatures are LTV (long term validatable) signatures, which are build iteratively in cascaded way by incorporating newer validation elements upon expiration of the ones collected before. Such signatures are expected to be validatable for long periods after the initial signature is created, even in case if one or more of the original CAs stop functioning. Advanced signatures are gaining extended popularity in European infrastructures, with a number of corresponding standards, such as PAdES [10], CAdES [11], XAdES [12], and ASiC [13] introduced and deployed lately on the governments' level.

**OCSP stapling in TLS.** OCSP stapling [14] is a technique used by TLS servers to reduce validation burden put on constrained TLS clients. Instead of pushing the client to performing a comprehensive certificate revocation check by connecting to external OCSP and CRL sources (which is assumed by normal TLS protocol flow), the server passes relevant OCSP responses (pre- fetched by it beforehand) to the client along with its certificate chain. This eliminates the need for the client to spend time and resources by requesting the certificate statuses from the online sources; it can verify the genuineness of the responses by checking the OCSP responders' digital signatures included in the responses.

## 4 Collision attack

We are going to show that the OCSP scheme as defined in [1] is subject to two important design flaws. Depending on configuration of the particular OCSP endpoint and on its role in the PKI environment, the exploitation of the vulnerabilities may result in an adversary obtaining the capability of generating forged OCSP responses signed with the OCSP responder's certificate, or even forged certificates if the OCSP responder shares its certificate and private key with the CA itself.

### 4.1 Predictability of responses

Let's get back to the OCSP protocol messages as shown on fig. 1 and 2. Consider a requestor making several consequent requests to the OCSP responder about the same certificate. Let's start with a simplified case where the requestor doesn't include any extensions to its request. Assuming that the status of the certificate doesn't change between the requests, the only elements in the *tbsResponseData* record (which is the object directly signed by the OCSP responder's certificate) that will change between the responses are time-related entries *producedAt*, *thisUpdate* and *nextUpdate*. The remaining elements of *tbsResponseData*, such as responder's ID and certificate identifiers, are fixed for a particular responder and certificate, and do not change between responses.

Given that the time values *producedAt*, *thisUpdate* and *nextUpdate* used by the responder for a request made at a particular moment in time can often be easily guessed by the requestor after black box research of the responder's behavior (primarily to establish server clock time bias and obtain a pattern of *thisUpdate* and *nextUpdate* evolution), the content of the *tbsResponseData* record generated by the responder is highly predictable by the requestor. In other words, basing on the knowledge on the responder's time management characteristics, the requestor can make an educated guess about the content of *tbsResponseData* record that is going to be generated and returned by the responder if they make a request at a specific moment in future.

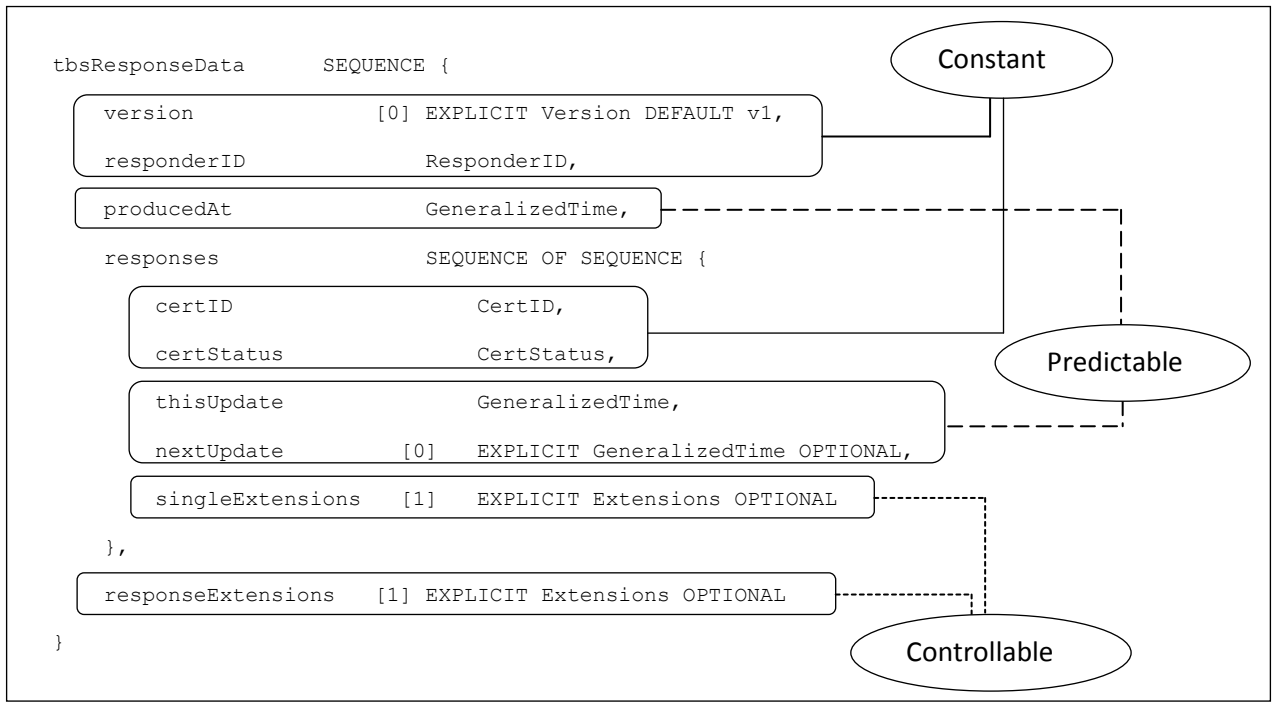

**Fig. 4.** Fields of *tbsResponseData* structure from the client's viewpoint.

Having gained some knowledge about the OCSP responder's setup, an adversary can build a local model of the responder in the form of interpolating function *TBSRESP*(*cert, time*), with *cert* being a certificate serial number and *time* a particular moment in future, which would be constructing *tbsResponseData* records that would match those returned by the responder if a status request for *cert* is submitted to it at the exact moment *time*.

As the responder always signs the contents it generates by applying its private key to the hash of *tbsResponseData* structure, the adversary that manages to construct a correct *TBSRESP*() implementation obtains the capability to pre-build a set of hashes that it can expect to be signed by the responder at any particular moment in future (for as long as the responder's configuration remains unchanged). To build the set, they simply run *TBSRESP*() in a loop over *cert* and *time* parameters with appropriate step values and within the needed scope:

$$hash_{ij} = H(TBSRESP(cert_i, t_j)), \ i = [0, |S|), \ t_j = t_0 + j\Delta t, \ j=[0, (t_1-t_0) / \Delta t) \ ,$$

where *H* is a hash function employed by the responder, *S* is a set of all valid certificate serial numbers (assume for simplicity that serial numbers are issued in series, from 0 to $|S|-1$); $t_0$ is the opening time moment, and $t_1$ is the closing time moment. The $\Delta t$ parameter specifies the responder's time granularity units, e.g. whole seconds. Each of the hashes $hash_{ij}$ could then be expected to match the hash calculated and signed by the responder if a properly formed request is submitted to it at the right moment $t_j$, $t_0 \leq t_j \leq t_1$.

The adversary can then use the set of hashes they have built to search for collisions. Note that no interaction with the OCSP responder is needed on this stage; the hashes can be constructed and the collisions searched completely autonomously. Once a collision is found, the attacker waits for the moment $t_j$ and submits the request, obtaining the responder's signature over the hash and discarding the *tbsResponseData* returned. For high-granularity or high-latency responders they may choose to use a high-frequency sequence of repetitive requests to increase the chances of triggering the responder at the right moment. The responder can therefore be viewed as an online signature generator for the set of pre-generated hashes.

The above considerations remain valid in a generalized case where the requestor and/or the responder choose to include any standard protocol extensions in their messages. As per [1], the only variable *tbsResponseData* extension is Nonce, whereas the others are fixed and do not change between the responses for long time. However, as the content of Nonce is effectively chosen by the requesting party (with the responder simply mirroring it back), it does not affect the adversary's capability to predict the content of the responses, and even extends it, as we will show below.

## 4.2 Scaling the attack: nonce- and serial number-based methods

While the set of hashes constructed with *TBSRESP*() function provides the adversary with important knowledge about the OCSP responder's future behaviour, this set is finite and capped with the length of the chosen time period $t_1$-$t_0$ and the cardinality of the serial number set $S$: the maximal number of hashes that can be predicted with *TBSRESP*() is $|S| \times ((t_1 - t_0) / \Delta t)$. For example, for the chosen time period of one day, the responder's time granularity of one second, and the set of valid serial numbers containing 1000 entries, the number of hashes that can be pre-generated sums up to $1000 \times 86,400 = 86,400,000$, which is not many in terms of collision search. Besides, the adversary might not want to wait for a distant moment in time to come (which may constitute weeks or even months) to launch the attack.

This is where Nonce extension comes handy. By adding the extension to its requests and varying the content of the nonce the adversary can scale the number of different *tbsResponseData* records they may expect to obtain from the responder at the same moment in time, effectively using the nonce as a random salt. The adversary can therefore define a target exploitation moment (or a narrow time period) $t_A$, and then pre-generate a set of hashes that could be signed at that particular moment by only altering the nonce value included in the request.

In other words, an attacker can build an advanced *TBSRESPEX*() function, by extending *TBSRESP*() with the nonce parameter:

$$hash_{ijk} = H(TBSRESPEX(cert_i, t_j, n_k)), i = [0, |S|], t_j = t_0 + j\Delta t, j=[0, (t_1-t_0) / \Delta t), k=[0, |N|) ,$$

where $N$ is a set of all possible nonce values.

The way in which a particular OCSP responder handles non-existent serial numbers can also play a role in scaling the attack. If the responder returns signed *unknown* status responses for non-existing certificates, fake serial numbers can be used as a salt instead of or on par with nonce to scale the number of hashes that can be pre-generated for a particular moment in time. It would be sufficient for the adversary to have the responder either support Nonce extension or respond with *unknown* response to non-existent certificate status requests to be able to attempt the collision attack.

The attack can be parallelized between any number of CPUs doing hash generation and collision search at the same time, and is easily scalable by choosing appropriate lengths for the nonce values. The cardinalities of well-formed nonce and serial number sets provide a powerful scalability instrument. For an *n*-byte nonce or serial number string, the number of hashes that can be pre-generated for a fixed moment in time is $8^n$. As the standards [1] and [2] do not impose any limitations on nonce or serial number lengths, an attacker can choose to use the values of appropriate length to generate the required number of hashes.

### 4.3 Birthday attack

Consider an adversary having a capability of building a different generator function *TBSFAKE*() that produces a series of random sources which the attacker would be interested in having signed by the responder's private key. *TBSFAKE*() can be built over existing *TBSRESPEX*() implementation if the attacker's goal is to obtain a falsified OCSP response for a particular certificate, or it can be implemented separately for sources of different nature.

The attacker then can start building two series of hashes, the first using *TBSRESPEX*() function and the second using *TBSFAKE*():

$$h_1^i = \text{H}(TBSRESPEX(cert_i, t_i, n_i)), cert_i \in S, t_i \in [t_0, t_1], n_i \in N \quad (1),$$

$$h_2^i = \text{H}(TBSFAKE(i)) \quad (2),$$

where $i = 1, 2, \ldots I_{max}$.

Any collision found between $h_1^u \in H_1$ and $h_2^v \in H_2$ presents a candidate for a successful attack, as it means that by sending a status request for certificate $cert_u$ at a moment $t_u$ with a nonce $n_u$ the attacker gets a signed hash matching the hash of the colliding input generated by $TBSFAKE(v)$.

In a particular scenario where the goal of the attacker is obtaining a falsified status for a particular certificate in the form of *BasicOCSPResponse* structure, they may use *TBSRESPEX*() as their *TBSFAKE*():

$$h_1^i = H(TBSRESPEX(cert_i, t_i, n_i)) , cert_i \in S, t_i \in [t_0, t_1], n_i \in N$$

$$h_2^j = H(TBSRESPEX(cert_{FAKE}, t_j, n_j)) , t_j \in [t_0, t_1], n_j \in N$$

where $i = 1, 2, \ldots I_{max}, j = 1, 2, \ldots, J_{max}$

As the inputs passed to function *H* on stages (1) and (2) have a random compound of arbitrary length, $H_1$ and $H_2$ can be viewed as having been built from a random source (where the fixed part of the input can be viewed as an additional initialization vector to the hash function). The attacker therefore can run a birthday attack to look for collisions between $H_1$ and $H_2$, reducing the complexity of finding a collision using brute force down to $1.25\ SQRT(image(H))$ on average [15].

### 4.4 Summary

To sum up the above, an attacker can build a completely autonomous, local application that pre-builds a database of hashes, each of which is known to be signed by the OCSP responder at any given moment in time in future. They can obtain a valid OCSP responder's signature over any of those hashes by crafting a properly formed status request and submitting it at the right moment. A random salt included in such requests in the form of a nonce extension or a serial number belonging to a non-existent certificate can be used to scale up the attack by generating any chosen number of valid hashes which could be expected to be signed by the responder at the particular moments in future.

OCSP responders have no possibility to find out that they have been/are being exploited, as all such status requests are valid and indistinguishable from legitimate ones. The lion's share of work necessary to facilitate the attack is performed by the attacker autonomously in advance without contacting the responder (they only need to conduct a few requests on the initial stage to learn about the responder configuration), so there is no way for the responders to realize that there is an attack being prepared against them.

The initial requests which the attacker needs to make to establish the fact of the responder's vulnerability to the attack and collect all the necessary configuration parameters, can be easily automated and performed without any involvement of the attacker. The attacker can therefore build a web crawler which would search for vulnerable responders across the web, or deploy a worm into a corporate network which would scan for vulnerable Intranet responders and report them back to the attacker.

## 5 Attack targets

This paragraph addresses the outcomes of a successful attack against an OCSP responder and their consequences for the victim. By a successful attack in this context we mean the discovery of a hash collision, followed by requesting a responder's signature over the colliding hash at the appropriate moment in time. The input to the hash function that produces a collision with one of pre-generated *tbsResponseData* variants may be of different types, which are be considered below.

**OCSP response forgery.** The breadth of the scope of sources that can be forged by obtaining a hash collision highly depends on the policies applied in the PKI environment the OCSP responder is a part of, and on the information contained in its certificate itself. If the responder uses a properly formed dedicated certificate (Fig. 2(b)), which typically restricts the purpose of the private key with OCSP response signing via the extended key usage id-kp-OCSPSigning flag, this scope is limited with OCSP responses originating from the same responder, as a source of any other kind signed with the responder's certificate won't pass the policy check. However, according to our research, a significant share of PKI environments don't stick to this requirement, by issuing OCSP certificates with id-kp-OCSPSigning flag omitted or accompanied with other irrelevant flags, such as id-kp-timeStamping, thus extending the applicability of the certificate. Other applications implement validation procedures improperly or use relaxed policy validation rules to provide compatibility with buggy implementations, by not validating certificate eligibility to sign content of a particular type. These purely technical peculiarities can be also taken into account when preparing the attack.

To generate a forged OCSP response for a particular certificate, the attacker runs the birthday attack over the two sets of outputs of *TBSRESPEX*() function as described above. The straightforward outcome of the attack is possession by the attacker of a forged certificate status record with pre-defined status and time values. That status record could then be used as a compound of more sophisticated attacks.

A minor fact which is worth mentioning here is that the attacker can employ the nonce-based scalability approach within their *TBSFAKE*() function independently of whether the nonce extension is actually supported by the responder.

**Certificate forgery.** A much more severe threat is posed by a scenario where an OCSP responder signs responses with the CA certificate itself (Fig. 2(a)), due to importance of the role of the CA private key in the PKI environment, and the breadth of the scope of its policy-regulated applicability.

In this case the attacker can use the responder to create a valid signature over a fake *TBSCertificate* record [2], effectively making the CA generate a valid X.509 certificate for them. The attacker has full control over the content of the fake *TBSCertificate* structure and can include any information they need. In particular, the forged certificate can be prepared to be a lower-level CA itself, by ticking the CA bit of its Basic Constraints extension, leading to the attacker being capable of creating its own PKI subtree that will chain up to the original trusted anchor. The attacker can also adjust revocation check-specific properties of the fake certificate in such way that the original CA will not be able to revoke it except by revoking the CA certificate itself.

Fake certificates generated in such way will be indistinguishable from legitimate ones, and it might take a while for the CA and PKI users to establish that a particular certificate is actually a fake. The problem of recognizing fake certificates is severely complicated by another OCSP design flaw. OCSP request and response formats only identify certificates by their serial numbers; no other information about the certificate (such as its key ID or hash) is exchanged. Therefore, if a forged certificate shares its serial number with a legitimate one, it would be impossible for the OCSP responder to tell whether a status request it received corresponds to the legitimate or a forged certificate.

The definition of the *TBSCertificate* structure, which provisions for a number of informational or flexibly variable elements, allows effortless construction of *TBSFAKE* that would generate a series of certificates with the pre-defined requirements (which include at least issuerRDN and authority key identifier fields matching the parameters of the CA/OCSP certificate) of arbitrary length. Such elements include, but are not limited to, an X.500 *uniqueIdentifier* field that is included as an entry in the TBSCertificate's *subject* field, the contents of subject key identifier extension, and any custom version 3 extension. As an attacker can pre-generate any needed quantity of *TBSCertificate* structures matching their criteria, they could employ the birthday attack mentioned above to look for collisions with *TBSRESPEX*().

All the above gives us the reason to state that the use of the CA certificate for the OCSP responder role should be considered an extremely dangerous practice and discouraged for compliant implementations.

**Attacks on higher-level protocols.** A potential capability of forging OCSP responses can be used in a variety of higher-level protocol attacks. A number of protocols use OCSP responses as standalone offline certificate status records and employ standard PKI procedures, like chain validation, to establish their integrity and level of trust.

One of such examples is OCSP stapling technique employed in TLS [14]. The technique aims to reduce the burden on TLS clients by including any OCSP responses needed to validate the TLS server's certificate chain into handshake messages sent by the server, effectively removing the need for the clients to contact the OCSP responders as part of the validation process. An attacker who gains access to the private key of a TLS server can continue to impersonate it indefinitely by providing fake OCSP responses. As long as such responses are valid and fresh enough, the clients will have no reason to contact the OCSP responder directly, and any changes in status of the TLS server certificate will remain unnoticed.

The other very widely used example is presented by long-term verifiable (LTV) signatures ecosystem [3]. The main idea of the long-term signatures is that they are required to be verifiable offline, without the need to contact any third-party PKI services (which might not exist at the time of signature validation in the distant future) directly. This is achieved by including all related validation information, such as CA certificates, CRLs and OCSP responses, into the signature, and updating it periodically to reflect changes in the PKI environment. As long-term signatures are mainly validated in offline mode, a fake response for a previously compromised certificate included in such signature could make detection of the fact of compromise extremely difficult.

The flexibility of specifying the lifetime of a particular OCSP response via *thisUpdate* and *nextUpdate* fields of the same allows the attacker to generate fake responses with reasonably long lifetimes to prevent the validating applications from requesting fresher responses online.

The ultimate outcome of the attack on the described OCSP vulnerability is creation or discouragement of a trust anchor, which can be used for altering normal validation procedures carried out by the prospective victims. The results of such alterations may vary and depend on the application layer protocols being attacked. Particular examples include invalidation of legal contracts and denial of non-repudiation (by intentionally signing the source with already revoked certificate), and intruding into access control procedures by impersonating a controlling TLS endpoint.

**Chosen plaintext attack on private key.** As an attacker has oracle access to the signing operation primitive, they can run a chosen plaintext attack [16] against the responder's private key or the signing algorithm.

While there are no effective chosen plaintext attacks on RSA and ECDSA known to date, there is no guarantee that such attack won't be discovered in future. It should be kept in mind that even if the attacker can't get use of the OCSP responder oracle to reveal the private key, they still have significant level of control over the signing process by being able to obtain valid signatures over a wide range of arbitrary plaintexts.

**Other considerations and implementation attacks.** The wide use of lightweight OCSP scheme may cause client-side implementations to employ more liberal processing of responses issued long time ago and remaining 'fresh' for a longer period of time. For example, a difference between *thisUpdate* and *nextUpdate* times in OCSP responses generated by Verisign's lightweight responder is one week, meaning that every particular response can be used within a week of issuance without contacting the responder for a fresher one. This may allow attackers to generate fake OCSP responses valid for longer periods, expanding the attack time span and increasing the chance of a successful attack.

As an OCSP responder can be used as an online signature generator, with loads of signatures being generated instantaneously on request, it can potentially be subject to various implementation attacks, such as RSA-CRT attack [17] or a timing attack [18]. In case where the OCSP responder's certificate is the CA certificate at the same time, it may pose an extra risk of the CA's private key exposure.

During our field experiments with live OCSP services we discovered another important implementation flaw. A number of OCSP responders appear to return Good status for serial numbers that were never issued. While not directly related to the topic covered by this paper, this peculiarity may be used as a component of a more sophisticated attack, including the one described in this paper.

# 6 Practical considerations

## 6.1 Attack costs

The success of the attack wholly depends on the susceptibility of the underlying hash function to collision attacks. In the worst scenario where no bespoke function-specific collision attacks are known, the attacker would have to resort to brute force birthday attack with an average number of tries of $2^{n/2}$ for a hash function that produces *n* bits of output.

Advances in computing power and availability of computing resources to general public through cloud services like EC2 made attacks relying on widescale computations technically accessible to everyone in the hacking community (provided that the attacker has enough funds to pay for the computing power and time), removing the long-existent capital investment barrier, and making computational attacks relatively cheap. There is a high probability that we can expect a spike in computational attacks which leverage public cloud services in the near future.

According to Bruce Schneier's 2012 survey [19], a full-on collision attack on SHA1 would cost $700K by 2015 and $173K by 2018. More recent research by Stevens, Karpman and Peyrin [20] provides significantly lower estimates for a SHA1 collision cost of between $75K and $120K over a few months of Amazon EC2 computing time. This price tag is well within the resources of a medium-size corporation or a criminal syndicate.

Despite its prospective retirement in 2017, SHA1 is still widely used worldwide, and there are expectations it will continue to be so for at least another few years. Even when SHA1 is fully decommissioned and replaced with SHA2 and SHA3, the attack will still pose a risk due to constant appearance of new attacks on more advanced hash algorithms, increasing availability of computing power, and its decreasing cost.

Note that the attacker may benefit from autonomous nature of the attack through optimizing/spreading the cost of the attack by increasing the duration of the attack and decreasing the burden on computational resources.

In order for an OCSP responder to be susceptible to the autonomous collision search attack, as well as to general known/chosen plaintext attacks on its private key, it should support and be configured to respond to nonce extension and/or provide meaningful responses for non-existent serial number requests. Support for any of these protocol features would allow the attacker to scale the attack to any extent they need.

**6.2 Real world observations**

To study the practical side of the problem, we implemented a software toolkit. The toolkit consists of two modules, with the first being capable of 'black box' assessment of a particular OCSP responder for its exposure to the flaws described above, and the second one emulating the attack by requesting the responder to sign a known hash at a given moment in time.

A set of public OCSP responders were picked for evaluation, including those of well-known CAs (Globalsign, Thawte, Verisign, Godaddy), EU Member States' national trust services, and solutions offered by private sector companies. In total 70 responders have been assessed, covering services originating from EU/EEA area (54), the United States (13), and few other countries (3).

The results obtained during the evaluation have shown that 53 (75.7%) of the services provide real-time certificate status information, allowing to make predictions about time values in distant future responses, with the other 17 (24.3%) sticking to the lightweight OCSP scheme and response caching. Of the services providing real-time information, 52 (74.3% of the whole set) support *tbsResponseData* content scaling either by mirroring the value of Nonce extension or by providing meaningful signed responses for non-existent certificates, thus exposing themselves for autonomous collision search attack.

The experiment also unveiled that as many as 40 services (57.1%) still use SHA-1 as their hashing algorithm, despite its use having been strictly discouraged by most industry standards and security and privacy regulations. Among those, 29 (41.4%) both use SHA-1 and support nonce- or non-existent certificate-based scaling, posing a high risk of a low-cost collision attack.

In total, 7 (10%) of the providers use their CA certificates for signing OCSP responses, with only one of them (1.4%) also supporting response scaling via either of the two methods. All providers use the granularity of one second for their time figures, making it simple for the prospective attackers to have their requests served at the needed moment in time.

Our experiment has also shown that as many as 22 (31.4%) responders return Good status for non-existent certificates. While being a separate issue to the one discussed in this paper, this feature still poses an important security flaw that should be identified and addressed.

**7 Suggestions and recommendations**

**7.1 Existing environments**

The primary tactics for dealing with the problem involves removing each of the weak points discussed above. We will consider them one by one below.

**Using dedicated certificates for OCSP services.** While not reducing the attacker's capabilities of running a successful hash pre-generation attack on the responder directly, the use of a dedicated certificate for an OCSP service removes the risk of generating fake certificates, which would look fully legit and verifiable within the PKI environment and indistinguishable from genuine. It is very important that such dedicated certificate is properly formed policy- and key usage-wise; in particular, the id-kp-OCSPSigning flag must be included and be the only flag in the certificate's extended key usage extension, and the basic constraints extension, if present, must have its CA field set to false.

**Increasing granularity of time values.** By using more granular time values the responder would significantly complicate the attacker's task of getting a response with desired *producedAt*, *thisUpdate* and *nextUpdate* entries to produce the pre-calculated *tbsResponseData* hash value. If a granularity of one second is used, the attacker can make a series of requests around the chosen time moment and expect that at least one of the returned responses would contain the time values they expect. Typically, they have a good chance to succeed with 20-50 requests made within a 5 seconds frame around the chosen time point. If the responder switches to one millisecond (0.001 second) granularity, it would be nearly impossible for the attacker to craft the request in time to get the response with desired time entries. By far, this is the simplest solution to employ.

**Using random biases for time values.** An alternative for the above solution is to apply random biases to the time values included in the response. Using a random 10 second or less bias for each time value (*producedAt*, *thisUpdate*, *nextUpdate*) is equivalent to increasing the granularity of any of the values up to 1 millisecond. Note that the random generator used for bias generation should be cryptographically strong to prevent the attacker from making guesses about the applied bias figures.

**Getting rid of the variable parameters.** Removing nonce support and implementing correct non-existent serial number handling procedures (by refusing to provide a signed response for serial numbers that were never used) significantly reduce the attacker's chances of running a successful attack, as they lose access to the scalability mechanisms. In this case they may only try a very limited set of message digests for every time moment, capped with the number of known valid certificate serial numbers in their possession.

At the same time, we admit that disabling nonce support on the responder side contradicts to widely accepted industry practices [21] and exposes the server and clients connecting to it to replay attacks. The decision whether to switch off nonce support should therefore be made with due care, and be preceded by in-depth risk assessment procedure.

If it is not possible to alter the OCSP implementation to handle non-existent serial numbers in a safe way, a viable alternative would be to use artificial random time values when handling such requests to eliminate the possibility of guessing the contents of the responses.

**Intelligent handling of nextUpdate.** As the attacker in their aim to pick a colliding response may try varying time figures, such as *thisUpdate* and *nextUpdate*, it makes sense to introduce a norm to the client-side validation policy to ignore responses with *nextUpdate* value exceeding certain threshold (such as an hour or a day), and always obtain a fresh response directly from the service if the existing one expires. The technique of client-side *nextUpdate* threshold might be useful not only in counteracting the described vulnerability, but also as an additional security measure against fake responses and buggy OCSP responders.

**Limiting the number of closely successive identical requests** would help to detect a possible attack in progress, with the attacker trying to obtain a response with pre-defined time figures.

### 7.2 Standard development

**Prohibiting the use of CA certificates for OCSP purposes.** We strongly believe in importance of discouraging the use of CA certificates for OCSP purposes on the requirements (standard) level. Our principal reason is a huge difference between levels of risk associated with the leakage of the CA and OCSP responder's private keys. Besides, OCSP responders, often residing at or even outside of the organisation's security perimeter, are more exposed to hostile Internet ecosystem, and as such are a much easier target for remote attacks. Separation of OCSP and CA services helps prevent exposure of the CA private key if an attack on the OCSP service succeeds.

**Employing responder-side nonce values.** A more substantial improvement to the OCSP standard would be introduction of a responder-side nonce extension. The idea is for responder to generate a new pseudorandom nonce for every response it returns independently of any request parameters. Provided that the PRNG used by the responder is cryptographically strong, the use of such extension would make prediction of the content of *tbsResponseData* nearly impossible.

Obviously, responders implementing the lightweight profile won't benefit from the server-side nonce extension, as the presence of a random entry in every response contradicts to the very idea of the profile. Still, as such responders are much less vulnerable to the guessing attack due to static nature of their responses, the impossibility to use the responder-side nonce technique is not vital for them.

**Using more specific certificate identifiers**. The fact that OCSP identifies certificates solely by their serial numbers makes it difficult for the responder to tell requests coming from legitimate applications from those intentionally constructed by an attacker. The current revision of the protocol provides no means for the responder to check whether the requestor actually has any knowledge about the certificate for which it is requesting the status, which makes the serial number-based attack scaling possible. The attacker can also learn the CA's serial number issuance pattern to make predictions about certificates that are going to be issued in near future, and use that knowledge for planning and/or optimising the collision attack.

Inclusion of more specific certificate information in the status request would make it easier for the responder to reject obviously fabricated requests and provide better service for legitimate ones. One of the options here is to include unique or nearly-unique certificate properties such as certificate or public key hash in the status request in addition to its serial number.

## 8 Conclusions

The vulnerabilities in the OCSP protocol described in the paper make it possible for an attacker to obtain a valid signature made with the responder's private key over arbitrary data.

While the attack requires a large amount of computing power, and is still impracticable for enthusiast hackers, the increasing availability of computational resources, both legitimate and exploitable, makes its cost go down steadily.

The autonomous nature of the attack and the attacker's ability to plan and exercise it in advance with minimal interaction with the victim make the attack nearly impossible to detect. The difficulties in obtaining any reliable evidence of the attack may affect the responder's and/or CA's non-repudiation commitments, and put the integrity of the corresponding PKI trust subtree at risk. The use of the same private key for purposes of way different grades may widen the surface of the attack and increase the risk to the environment drastically.

The paper suggests appropriate countermeasures to be taken on the policy, protocol and implementation levels. The recommended changes to the OCSP protocol rely on the protocol's extension mechanism and are non-breaking. The flaws pointed out in the paper go beyond the scope of the OCSP service, and should be taken into account when designing new security protocols.